\documentclass{PoS}
\usepackage{todonotes}

\title{Relic neutrinos: local clustering and consequences for direct detection}

\ShortTitle{Relic neutrino clustering}

\author{\speaker{Stefano Gariazzo}\\
        IFIC (CSIC / University of Valencia)\\
        E-mail: \email{gariazzo@ific.uv.es}}

\abstract{
The Cosmic Neutrino Background is a prediction of the standard cosmological model, but it has been never observed directly. Experiments with the aim of detecting relic CNB neutrinos are under development. For such experiments, the expected event rate depends on the local number density of relic neutrinos. Since massive neutrinos can be attracted by the gravitational potential of our galaxy and cluster locally, a local overdensity of relic neutrinos should exist at Earth. We report the status of our knowledge on the clustering of neutrinos and the consequences for future direct detection experiments.
}

\FullConference{
European Physical Society Conference on High Energy Physics - EPS-HEP2019 -\\
			10-17 July, 2019\\
			Ghent, Belgium}

\newcommand{\e}[1]{\ensuremath{\times10^{#1}}}
\newcommand{\Neff}{\ensuremath{N_{\rm eff}}}
\newcommand{\eqref}[1]{(\ref{#1})}

\begin{document}

\section{Relic neutrino background}
The standard cosmological model predicts a number of relics that were produced during the initial phases of the Universe evolution.
While the Cosmic Microwave Background (CMB) and the light elements produced during the Big Bang Nucleosynthesis (BBN)
have already been observed,
we still miss a direct confirmation of the Cosmic Neutrino Background (CNB),
a relic of neutrinos which decoupled from the thermal plasma when the temperature of the Universe
was around the MeV scale and the Universe approximately 1~s old (see e.g.~\cite{Lesgourgues:2018ncw}).
Such neutrinos, after freely streaming during more than 13 billion years,
have now a mean energy of $\sim5\e{-4}$~eV and an average number density of $n_0\sim56$~cm$^{-3}$
per family and per neutrino/antineutrino state.

We have only an indirect probe of their existence,
which comes from the fact that they were ultra-relativistic when decoupling.
One can write the energy density of radiation in the early universe
by means of the effective number of relativistic degrees of freedom,
\begin{equation}\label{eq:Neff}
\Neff
=
\frac{8}{7}
\left(\frac{11}{4}\right)^{4/3}
\frac{\rho_\nu}{\rho_\gamma}\,,
\end{equation}
where $\rho_\gamma$ and $\rho_\nu$ represent the energy density of photons and neutrinos, respectively.
The expected value of \Neff, which is $\sim3.04$ when considering the standard three-neutrino oscillations
in the early Universe \cite{Mangano:2005cc,deSalas:2016ztq,Gariazzo:2019gyi},
is confirmed by experimental determination by the observation of the CMB anisotropies
($\Neff=2.92^{+0.36}_{-0.37}$ at 95\% CL when considering the full temperature and polarization spectra of the CMB \cite{Aghanim:2018eyx})
and
BBN abundances
($\Neff=2.90\pm0.22$ at 68\% CL \cite{Peimbert:2016bdg}).
Despite this is a very strong indirect confirmation of the existence of three neutrino-like particles
in the early Universe,
only a direct detection of the CNB would ensure us that such particles are really the standard model neutrinos.

In order to directly detect the relic neutrinos, however,
one must consider a process that requires no energy threshold.
One of the most promising methods is
neutrino capture on beta-decaying nuclei \cite{Weinberg:1962zza}.
If one considers a nucleous that can undergo $\beta$-decay ($n\rightarrow p+e^-+\bar\nu$),
a relic neutrino can trigger the emission of an electron through the reaction
\begin{equation}\label{eq:nucapture}
n+\nu\rightarrow p+e^-\,.
\end{equation}
Such electron will have an energy larger than the $Q$-value available in the reaction:
the capture of relic neutrinos would therefore generate a peak
located just above the end point of the $\beta$-decay spectrum,
which turns out to be the main background for such search.
In order to distinguish the signal from the background events, therefore,
one needs an energy resolution of the same order of the neutrino mass or better \cite{Long:2014zva}.
For experimental purposes
(in order to limit the contamination from $\beta$-decay events and have a reasonably high cross-section),
the best candidate for performing such search turns out to be tritium \cite{Cocco:2007za}.

\section{The PTOLEMY project}
Based on the idea of neutrino capture,
for the first time an experiment which aims at the detection of relic neutrinos is under development.
Such project is named PTOLEMY%
\footnote{\href{https://ptolemy.lngs.infn.it}{https://ptolemy.lngs.infn.it}.}
\cite{Betts:2013uya,Baracchini:2018wwj}
and plans to have a final configuration with 100~g of tritium,
which would ensure at least few signal events per year,
and an energy resolution of $50-100$~meV.
In order to reach these targets,
seriously challenging experimental efforts will be needed.
The precise setup of PTOLEMY will be defined in the incoming years,
tough many studies are in progress (see e.g.~\cite{Betti:2018bjv}).

The physics case of PTOLEMY,
and more in general of experiments which aim at a direct detection of the CNB
with the neutrino capture method,
has been discussed in Ref.~\cite{Betti:2019ouf}.
One of the most crucial aspects is the determination of the signal event rate.
This can be written as:
\begin{equation}\label{eq:gammacnb}
\Gamma_{\rm CNB}
=
N_T
\bar\sigma
v_\nu
\sum_{i=1}^{N_\nu}
|U_{ei}|^2
n_i
\,,
\end{equation}
where
$N_T$ is the number of tritium atoms in the source,
$\bar\sigma$ is the average cross section for neutrino capture,
$v_\nu$ is the neutrino velocity,
$N_\nu$ is the number of neutrino mass eigenstates,
$|U_{ei}|^2$ represents the mixing of the $i$-th neutrino mass eigenstate with the electron flavor eigenstate
and
$n_i$ is the local number density of neutrinos that can interact.
The total $\Gamma_{\rm CNB}$ is expected to be $\sim4$~yr$^{-1}$
when considering Dirac neutrinos and 100~g of tritium \cite{Long:2014zva}.
If the energy resolution is sufficiently small, one can in principle distinguish the various peaks
in the electron spectrum, that correspond to the capture of each neutrino mass eigenstate.
The amplitude of the various peaks depends on $|U_{ei}|^2$, as one can see in Eq.~\eqref{eq:gammacnb},
and therefore on the mass ordering.

\begin{figure}[t]
\includegraphics[width=0.5\textwidth]{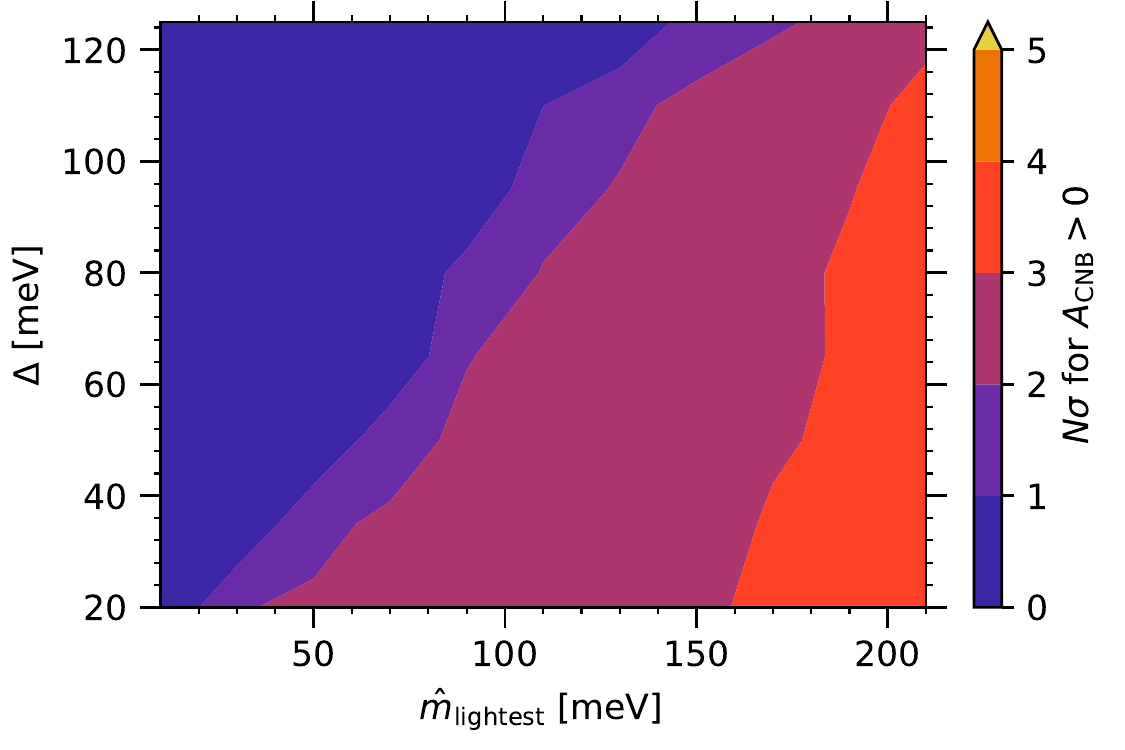}
\includegraphics[width=0.5\textwidth]{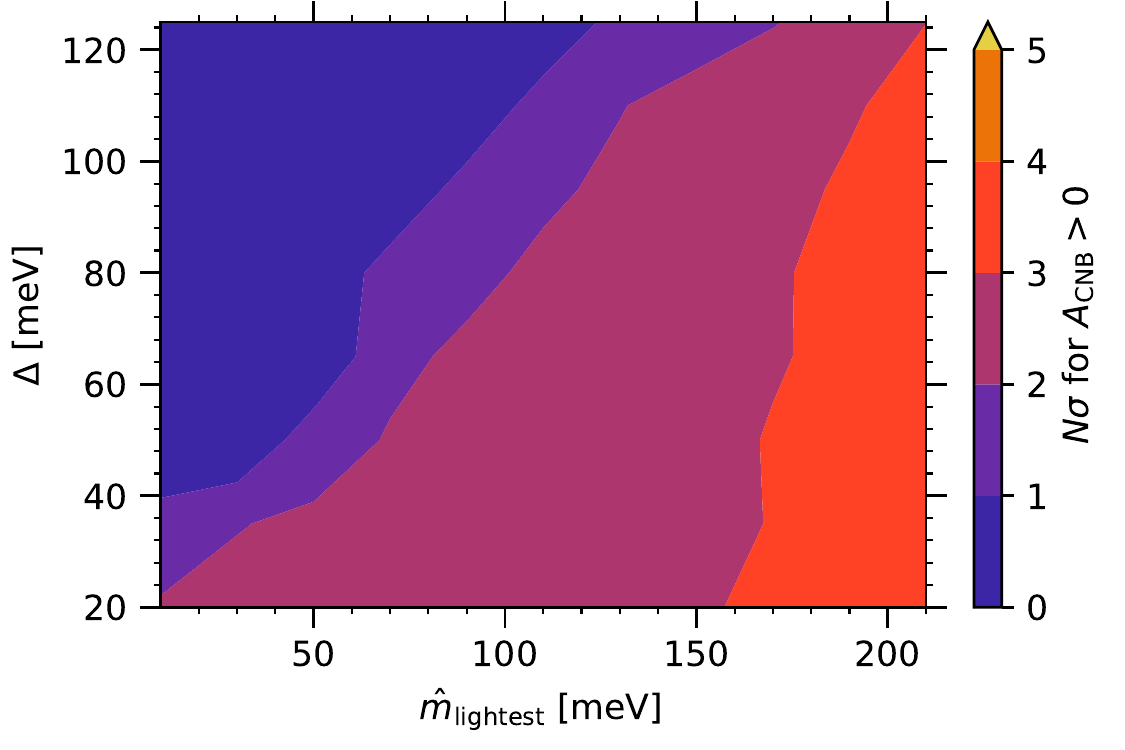}
\caption{\label{fig:pto_sens}
Significance for the direct detection of relic neutrinos in the full-scale PTOLEMY experiment,
considering normal (left) or inverted (right) ordering,
as a function of the lightest neutrino mass and the energy resolution.
From \cite{Betti:2019ouf}.
}
\end{figure}

Such numbers would allow a direct detection of the CNB
only if the lightest neutrino mass is larger than the energy resolution,
as one can see in Fig.~\ref{fig:pto_sens}, from \cite{Betti:2019ouf}.
In the case of inverted ordering, the detection is somewhat easier for small neutrino masses,
as the main peak of the CNB capture correspond to the heaviest mass eigenstates,
whose mass cannot be smaller than $\sim50$~meV.

One interesting note is that the number of interacting states,
if neutrinos are non-relativistic, depends on their nature.
While if neutrinos are Dirac particles only left-handed neutrinos can participate the reaction in Eq.~\eqref{eq:nucapture},
in the Majorana case, neutrinos with both chiralities can interact.
As a consequence, the event rate may be twice as large in the Majorana than in the Dirac case \cite{Roulet:2018fyh}.
Another possible increase of the event rate comes from the fact that non-relativistic neutrinos may have fallen
in the gravitational potential of our galaxy, so that their local number density
can be larger than the average one.

\section{Clustering of relic neutrinos}
The clustering of relic neutrinos in the gravitational potential of the Milky Way (MW) or of any larger object,
in the past,
has been studied with different methods.
$N$-body simulations are suitable to compute the shape of the neutrino halo in large objects,
such as galaxy clusters \cite{VillaescusaNavarro:2011ys},
but may have troubles in resolving the small structures, galaxies for example.
The authors of Ref.~\cite{Singh:2002de} overcame this problem
using a method based on the resolution of the collisionless Boltzmann equation,
which assumes that the dark matter evolves independently of the presence of neutrinos.

Another method that has been proposed is named ``$N$-one-body'' simulation \cite{Ringwald:2004np},
and consists in evolving $N$ test particles in the gravitational potential of dark matter and baryons,
which again is assumed to be unaffected by neutrinos.
Moreover, each neutrino is assumed to be independent of the others
and the number of test particles must be high enough to have an adequate sampling
of the available initial positions and momenta.
In order to keep the number of particles under control, this method needs to be used within the assumption
of spherical symmetry, for which only 3 directions in the parameter space (1 for the distance, 2 for the momentum)
need to be sampled.
Once the final position of each test particle is computed,
one can obtain the profile of the neutrino halo by smearing each test particle with a selected kernel
(a Gaussian one, for example)
and weighing each one according to its corresponding initial parameter space.
Since the evolution equations for the neutrinos in the gravitational potential can be written
independently of the neutrino mass,
one can compute a set of $N$ test particles with a given mass and
then rescale the entire simulation for a different mass,
re-weighing all the test particles \cite{Ringwald:2004np,Zhang:2017ljh}.

The first results \cite{Ringwald:2004np}
computed with the $N$-one-body method were revised in Ref.~\cite{deSalas:2017wtt},
where an updated description of the MW content was adopted.
In particular, the considered dark matter profiles were updated to take into account
some of the most recent astrophysical observations \cite{Pato:2015tja},
while for the baryons both new astrophysical data \cite{Misiriotis:2006qq}
and results from $N$-body simulations \cite{Marinacci:2013mha} have been considered.
The obtained clustering factors, defined as the local neutrino density divided by the average one,
were in the range 1.1--1.2 for neutrinos of 60~meV mass
and 1.7--2.9 for neutrinos with a mass of 150~meV, depending on the astrophysical setup.

The MW, however, is not the only object in the Universe.
Results from full $N$-body simulations \cite{VillaescusaNavarro:2011ys}, indeed,
pointed out that galaxy clusters may have a neutrino halo that extends to distances above 10~Mpc,
and we know that the Virgo cluster is located at about 16~Mpc from us
(see e.g.~the NASA Extragalactic Database%
\footnote{\href{https://ned.ipac.caltech.edu/}{https://ned.ipac.caltech.edu/}.}).
Also, the results of \cite{deSalas:2017wtt} demonstrated that the neutrino halo of a galaxy like ours
can extend to $\sim1$~Mpc, approximately the distance of the Andromeda galaxy \cite{Kafle:2018amm}.
A more complete calculation, therefore,
needs to take into account a more complex astrophysical environment,
for which relaxing the assumption of spherical symmetry is required.

The $N$-one-body method as explained above
can serve as basis for a different calculation.
Since we are only interested at the value of the neutrino number density at Earth,
all the test particles that end up far from us at $z=0$ are not necessary to obtain the final result.
A smart idea is to select the $N$ test particles only among those that arrive at Earth today,
and evolve their trajectories backwards in order to determine
their initial positions and momenta at early times.
The conservation of phase-space density along the trajectories (the Liouville's theorem)
guarantees that it is possible to weigh the neutrinos a posteriori
according to their initial (homogeneous and isotropic) phase space volume
and consequently obtain the clustering factor at Earth.

Such method has been recently tested in Ref.~\cite{Mertsch:2019qjv},
where a new calculation of the clustering factor at Earth as been proposed.
For the first time, in Ref.~\cite{Mertsch:2019qjv}
a ``backwards $N$-one-body'' method has been adopted to compute the local relic neutrino density
without assuming spherical symmetry.
Concerning our galaxy, the description of the dark matter is similar
to the one used previously \cite{deSalas:2017wtt,Zhang:2017ljh},
while the baryon contribution has been computed in cylindrical symmetry
instead of approximating the distribution as spherical.
Moreover, additional to the MW the authors considered the contribution
of the Andromeda galaxy and of the Virgo cluster.
Concerning the former, only dark-matter has been taken into account,
as the exact shape of the baryon profile is not expected to play a significant role,
given the distance between Andromeda and the MW (see below).

\begin{figure}[t]
\includegraphics[width=0.5\textwidth,clip,viewport=26 17 320 272]{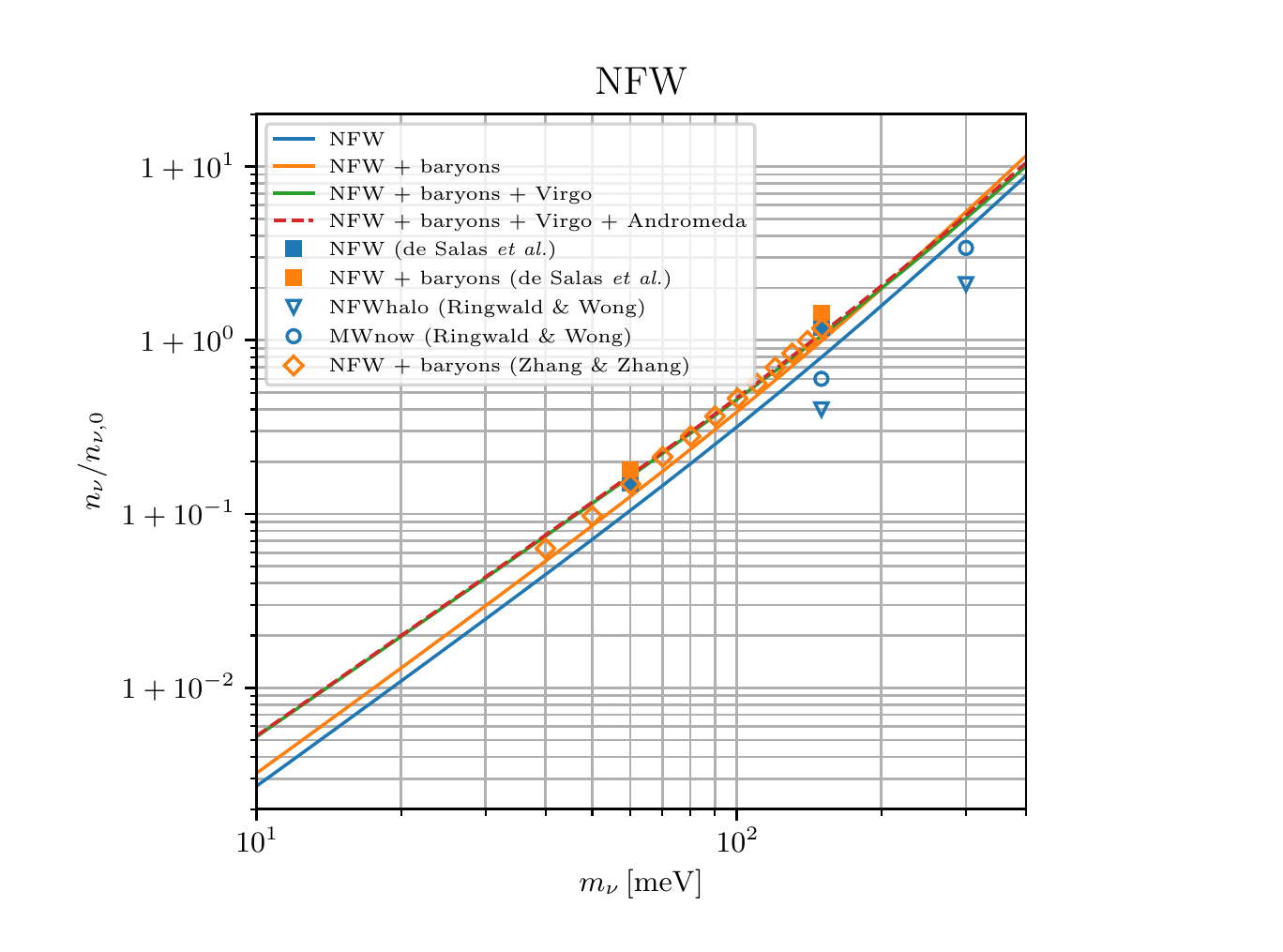}
\includegraphics[width=0.5\textwidth,clip,viewport=26 17 320 272]{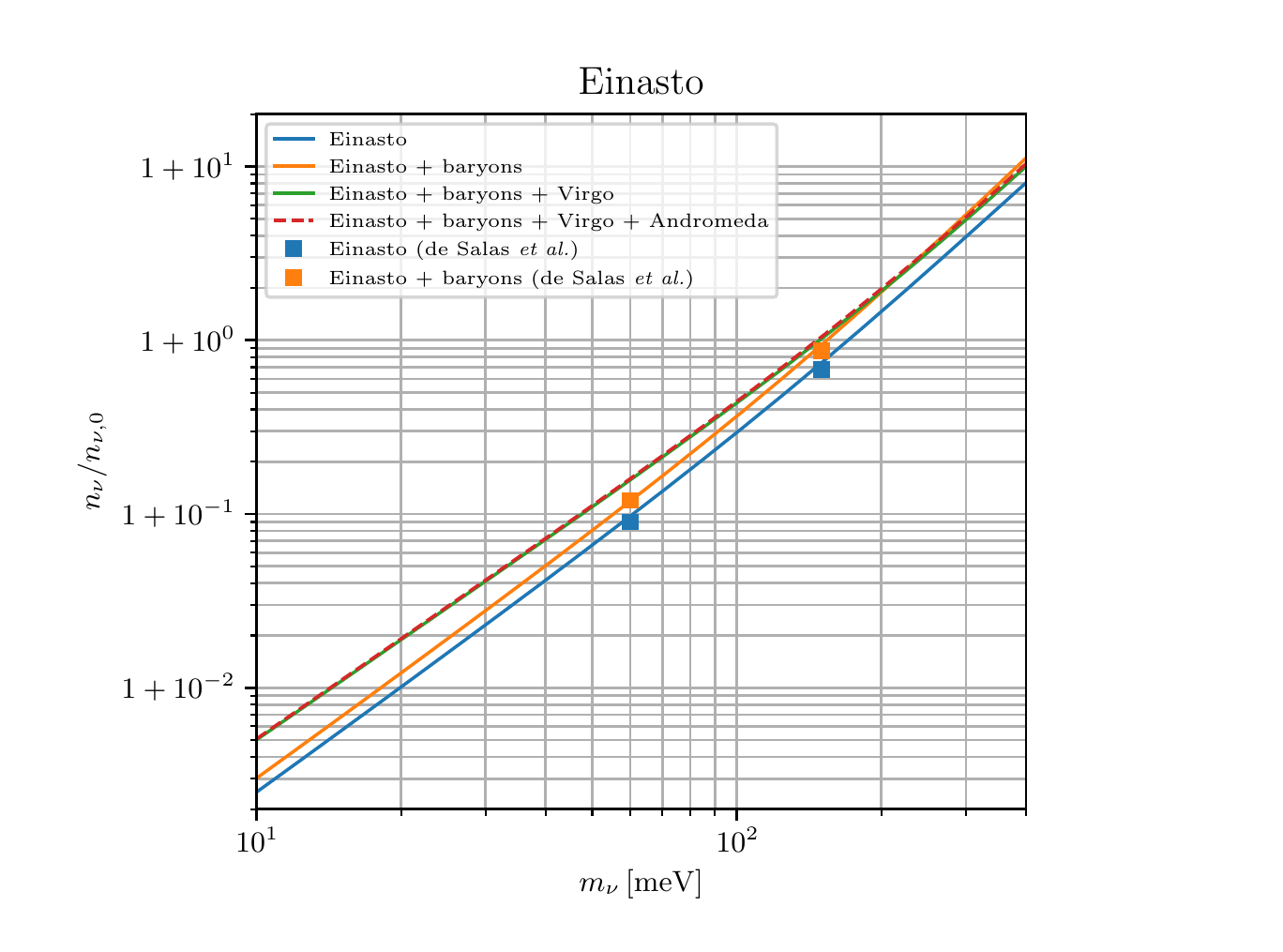}
\caption{\label{fig:nfw_cf}
The local relic neutrino density as a function of the neutrino mass,
considering different assumptions on the local astrophysical environment,
in comparison with the results from previous papers
(\cite{Ringwald:2004np,deSalas:2017wtt,Zhang:2017ljh}).
From \cite{Mertsch:2019qjv}.
}
\end{figure}

The results of Ref.~\cite{Mertsch:2019qjv} are summarized in Fig.~\ref{fig:nfw_cf},
when considering a NFW (left panel) or Einasto (right panel) profile for the dark matter in the MW.
The solid and dashed curves represent the results obtained in the paper,
using different astrophysical configurations,
while the various points show the results of previous calculations
(triangles and circles are from \cite{Ringwald:2004np},
squares from \cite{deSalas:2017wtt}
and diamonds from \cite{Zhang:2017ljh}).
Note that the NFW description of \cite{Mertsch:2019qjv}
is slightly different from the one in \cite{deSalas:2017wtt,Zhang:2017ljh}.

As the figure shows, the clustering factor has a clear dependence on the neutrino mass:
heavier (and slower) neutrinos cluster more than lighter ones.
It is also interesting to note that the Virgo cluster contributes to increase significantly $n_\nu$
when the neutrino mass is very small, as its neutrino halo is more extended,
while for high neutrino masses it can have the opposite effect:
since the total number of neutrinos is always the same,
when the Virgo cluster attracts more of them,
the number of neutrinos that can cluster around the MW decreases.
As a consequence, the local number density of relic neutrinos grows less
when the Virgo cluster is included than when it is not considered in the calculation.
Finally, we can also notice that Andromeda contributes only slightly to the amount of clustering at Earth,
at the point that the green line (MW+Virgo)
is almost indistiguishable from the red one (MW+Virgo+Andromeda).
Given the extremely small contribution from Andromeda, moreover,
we conclude that it is safe to ignore the presence of smaller objects as the MW satellite galaxies,
because their mass is much smaller than the one of the MW.

\section{Conclusions}
The existence of the cosmic neutrino background has been confirmed indirectly,
but a direct detection of the relic neutrinos is envisaged in order to confirm
that such background exists and has the expected properties.
A direct detection through the process of neutrino capture on $\beta$-decaying nuclei,
adopted by the PTOLEMY project,
will open the road for exciting studies,
such as the first observation of non-relativistic neutrinos,
a measure of their mass and a confirmation of the cosmological model.
The goal is however very ambitious.
According to recent estimates \cite{Betti:2019ouf},
the expected energy resolution ($\sim50-100$~meV) and tritium amount (100~g in the full configuration)
will allow an observation of the relic neutrinos only if the mass of the lightest neutrino is not too small.
The fact that neutrinos can cluster in the local gravitational potential enhances
the expected signal event rate, but for realistic neutrino masses the gain is small.
Since the crucial point remains to distinguish the relic neutrino events from the $\beta$-decay background,
a small enhancement in the event rate does not significantly increase the detection possibilities,
if the energy resolution is not sufficient,
but the increased event rate would make easier
an observation of the interesting events over external (non $\beta$-decay) backgrounds.
Moreover, a precise determination of the event rate will open the possibility to constrain
not only the Dirac-Majorana nature of neutrinos,
but also the composition of the astrophysical environment where the relic neutrinos can cluster.

\section*{Acknowledgements}
The author receives support from the European Union's Horizon 2020 research and innovation programme under the Marie Sk{\l}odowska-Curie individual grant agreement No.\ 796941.

\bibliography{main}

\end{document}